# Orbits and emission spectra from the 2014 Camelopardalids

José M. Madiedo[1, 2], Josep M. Trigo-Rodríguez[3], Jaime Zamorano[4], Jaime Izquierdo[4], Alejandro Sánchez de Miguel[4], Francisco Ocaña[4], José L. Ortiz[5], Francisco Espartero[1, 5, 6], Lorenzo G. Morillas[7], David Cardeñosa[8], Manuel Moreno-Ibáñez[3] and Marta Urzáiz[3].

[1] Facultad de Física, Universidad de Sevilla, Departamento de Física Atómica, Molecular y Nuclear, 41012 Sevilla, Spain
[2] Facultad de Ciencias Experimentales, Universidad de Huelva. 21071 Huelva (Spain).
[3] Institut de Ciències de l'Espai (CSIC-IEEC), Campus UAB, Facultat de Ciències, Torre C5-parell-2ª, 08193 Bellaterra, Barcelona, Spain.
[4] Depto. de Astrofísica y CC. de la Atmósfera, Facultad de Ciencias Físicas, Universidad Complutense de Madrid, 28040 Madrid, Spain.
[5] Instituto de Astrofísica de Andalucía, CSIC, Apt. 3004, Camino Bajo de Huetor 50, 18080 Granada, Spain.
[6] Observatorio Astronómico de Andalucía, 23688 La Pedriza, Alcalá la Real, Jaén, Spain.
[7] IES Las Fuentezuelas, Avd. Arjona 5, 23006 Jaén, Spain.
[8] La Fecha Observatory, MPC J34, Valladolid, Spain.

**ABSTRACT**

We have analyzed the meteor activity associated with meteoroids of fresh dust trails of Comet 209P/LINEAR, which produced an outburst of the Camelopardalid meteor shower (IAU code #451, CAM) in May 2014. With this aim, we have employed an array of high-sensitivity CCD video devices and spectrographs deployed at 10 meteor observing stations in Spain in the framework of the Spanish Meteor Network (SPMN). Additional meteoroid flux data were obtained by means of two forward-scatter radio systems. The observed peak zenithal hourly rate (ZHR) was much lower than expected, of around 20 meteors $h^{-1}$. Despite of the small meteor flux in the optical range, we have obtained precise atmospheric trajectory, radiant and orbital information for 11 meteor and fireball events associated with this stream. The ablation behaviour and low tensile strength calculated for these particles reveal that Camelopardalid meteoroids are very fragile, mostly pristine aggregates with strength similar to that of the Orionids and the Leonids. The mineral grains seem to be glued together by a volatile phase. We also present and discuss two unique emission spectra produced by two Camelopardalid bright meteors. These suggest a non-chondritic nature for these particles, which exhibit Fe depletion in their composition.

# 1 INTRODUCTION

Jenniskens (2006) first suggested the possibility of meteor activity associated with 209P/LINEAR on 24 May 2014, with some of the dust trails produced by this comet between 1798 and 1979 passing at a really close distance of about 0.0002 AU from Earth around 7h UT. Subsequent studies confirmed that comet 209P/LINEAR's 18th, 19th and some 20th century dust trails would be in Earth's path on 24 May 2014 between 6:00 and 8:00 UT, predicting a meteor shower radiating from the constellation of Camelopardalis (Vaubaillon 2012, Jenniskens and Lyytinen 2014). Assuming that the comet was weakly active also in those years, peak ZHR values of 100-400 were predicted (Vaubaillon et al., 2012, Maslov 2013, Ye et al., 2014). If the comet was more active in the past, then rates might be higher (Rao, 2013). The possibility of a meteor storm with over 1000 meteors per hour attracted much attention from the public and the media. But of course also from meteor and comet scientists, since the observation of this outburst of the Camelopardalid meteor shower could provide key information about the dynamical past and dust production of 209P/LINEAR.

Despite observers in North America were better positioned to monitor the Camelopardalids during the predicted activity peak, Europe also offered good opportunities to spot some of these meteor events, especially from the west of this continent. With this aim, CCD video observations of the Camelopardalids were performed from the SPMN meteor stations along Spain. In addition to this optical monitoring, two forward-scatter radio systems were used to analyze the activity of this meteor shower. Since these systems work in a continuous way, our monitoring was not limited to the predicted maximum activity date. As a consequence of this, the behaviour of the Camelopardalids was also characterized during several days before and after the activity peak. We present here, for the first time, orbital elements and emission spectra of optical meteors associated with these dust trails and demonstrating the pristine nature of 209P/LINEAR forming materials.

# 2 INSTRUMENTATION AND METHODS

In order to measure meteoroid trajectories and pre-atmospheric orbits by triangulation, low-light video cameras (models 902H and 902H Ultimate from Watec Co., Japan) operating at sites listed in Table 1 were employed. A detailed description of these systems has been published elsewhere (Madiedo & Trigo-Rodríguez 2008; Madiedo et al. 2010). For data reduction we have employed the AMALTHEA software (Madiedo et al. 2011, 2013b), which was developed by the first author and follows the methods described in Ceplecha (1987).

Meteor emission spectra were obtained by attaching holographic diffraction gratings (500 or 1000 lines/mm, depending on the device) to the objective lens of some of the above-mentioned CCD video cameras. With these videospectrographs we can record the emission spectrum of meteors brighter than magnitude -3/-4 (Madiedo et al. 2013a; Madiedo 2014).

In order to measure rates, two forward-scatter radio systems operating at a frequency of 143.05 MHz were used. These were located at Jaén (South of Spain) and Valladolid (Northwest of Spain). The first of these systems employed an 8 dBi 6-elements yagi antenna and a Yaesu FT817 ND radio receiver. At Valladolid, the observations were carried out with a RTLSDR R820T radio receiver connected to a 2 dBi discone-type antenna. Both forward-scatter systems received the reflections from the GRAVES radar, located in Dijon (France).

## 3 RESULTS

As shown in Figure 1, the forward scatter systems revealed that the peak activity of the Camelopardalids took place on May 24th between 6:00 and 8:00 UTC, as predicted (Jenniskens 2006, Vaubaillon 2012, Maslov 2013, Jenniskens and Lyytinen 2014, Ye et al. 2014). At this peak, the Canadian Meteor Orbit Radar (CMOR) detected a meteoroid flux comparable with the 2011 October Draconid outburst (Brown, pers. comm.). In spite of this, the shower in the optical range was disappointing, with a peak ZHR according to the International Meteor Organization of about 20 meteors $h^{-1}$ (www.imo.net). Thus, Brown (2014) found that shower radar echoes recorded by CMOR were confined to faint meteors (equivalent visual magnitude 6-7), consistent with the debris trails populated mainly by particles of milligram mass and smaller. The low activity observed in the optical range could be a consequence of the low dust-production activity found for Comet 209P, and might support the idea that this object is currently transitioning to a dormant comet (Ye et al. 2014). Our analysis of the particles' tensile strength, which is presented in the following sections, reveals another possible cause for the low flux of meteors observed in the optical range that will be proposed below. In total, we imaged 15 double-station Camelopardalids from May 22 to May 28. However, just 11 of these provided good quality atmospheric trajectories and orbits, as explained below. These 11 meteors are listed in Table 2, where their SPMN codes are also indicated for identification. The brightest of these were events SPMN240514C (abs. mag. -5.0 ± 0.5) and SPMN260514A (abs. mag. -8.0 ± 0.5). The latter was recorded 2 days after the peak activity date. The lightcurve of these two Camelopardalids are shown in Figure 2. As can be noticed, the lightcurve of meteor SPMN240514C is rather smooth, but SPMN260514A exhibited three main flares during the second half of its atmospheric path. These took place at times $t_1$ = 1.82 s, $t_2$ = 1.98 s and $t_3$ = 2.34 s. During these flares the meteor reached an absolute of -4.1 ± 0.5, -8.0 ± 0.5 and -1.9 ± 0.5, respectively.

### 3.1 Orbital elements

We have employed the planes intersection method (Ceplecha 1987) to obtain the atmospheric trajectory and radiant position of our double-station Camelopardalids. In our analysis, we have considered only those events for which the convergence angle is above 20º. This is the angle between the two planes delimited by the observing sites and the meteor path in the triangulation, and measures the quality of the determination of the atmospheric trajectory (Ceplecha 1987). For the 11 multi-station meteors satisfying this condition, Table 2 summarizes the calculated values of the beginning and terminal heights ($H_b$ and $H_e$, respectively), the preatmospheric, geocentric and heliocentric velocities ($V_\infty$, $V_g$ and $V_h$, respectively), the equatorial coordinates (J2000.0) of the geocentric radiant ($\alpha_g$ and $\delta_g$), and the absolute peak magnitude (M). The standard method described in Ceplecha (1987) provided the heliocentric orbit of the progenitor meteoroids (Table 3). The calculated position of the geocentric radiant for each meteor is shown in Figure 3.

### 3.2 Tensile strength

Most of these meteors exhibited a quasi-continuous ablation behaviour, with very smooth lightcurves, but other events were observed to exhibit at least one flare along their atmospheric path. This is the case for the double-station meteors SPMN240514C, SPMN260514A and SPMN280514A. These flares have been employed to estimate the tensile strength of these meteoroids by following the usual procedure (Trigo-Rodríguez & Llorca 2006). Thus, this parameter has been estimated as the aerodynamic pressure P at the disruption point:

$$P = \rho_{atm} \cdot v^2 \qquad (1)$$

where v and $\rho_{atm}$ are the velocity of the meteoroid and the atmospheric density at the height where this fracture takes place, respectively. We have calculated this density by using the US standard atmosphere model (U.S. Standard Atmosphere 1976). According to this, the strength of these meteoroids yields $(2.9 \pm 0.4) \cdot 10^4$ dyn cm$^{-2}$ (for SPMN240514C), $(9.0 \pm 0.4) \cdot 10^4$ dyn cm$^{-2}$ (for SPMN260514A) and $(8.3 \pm 1.2) \cdot 10^3$ dyn cm$^{-2}$ (for SPMN280514A), with an average value of $(4.2 \pm 0.3) \cdot 10^4$ dyn cm$^{-2}$.

### 3.3 Emission spectra

The emission spectra associated with events SPMN220514B and SPMN260514A were recorded from the meteor stations at La Pedriza and Sevilla, respectively. However, since the first of these signals was very dim it has provided very limited information. Thus, this signal only shows the emission corresponding to the Na doublet at 588.9 nm and the Mg triplet at 516.7 nm. Fortunately the spectrum produced by event SPMN260514A showed more contributions. The calibration in wavelength of these spectra and the identification of the main contributions have been performed with the CHIMET software (Madiedo et al. 2013b). The processed spectrum of the SPMN260514A event, once corrected by taking into account the spectral sensitivity of the recording device, is shown in Figure 4a. In this plot, multiplet numbers are given according to Moore (1945). The most significant contributions correspond to the Mg I-2 triplet at 516.7 nm and the Na I-1 doublet at 588.9 nm, but also to the combined emissions from Fe I-4 (385.6 nm) and Mg I-3 (383.8 nm), which appear blended. The contributions from multiplets Fe I-23 (358.1 nm), Fe I-5 (373.7 nm), Ca I-2 (422.6 nm), Fe I-41 (441.5 nm) and Fe I-15 (532.8 nm) were also identified. The calibrated spectrum of event SPMN220514B is plotted in Figure 4b, where the emissions from Mg I-2 (516.7 nm) and Na I-1 (588.9 nm) are shown.

## 4 DISCUSSION

Meteors recorded during the night from May 23 to May 24 were most likely produced by the fresh dust trails ejected from 209P between 1798 and 1979, while those imaged outside that period would correspond to the activity background of the Camelopardalid shower resulting from older dust trails produced by this comet. Figure 3 clearly shows that the dispersion in the geocentric radiant is higher for meteors belonging to the background component. The averaged coordinates of the geocentric radiant obtained from the analysis of the 7 Camelopardalids imaged during the main activity period yield $\alpha_g = 121.9 \pm 1.1°$, $\delta_g = 78.3 \pm 0.4°$, which fits fairly well the predicted value (Ye et al. 2014) and the position inferred from radar observations (Brown 2014) ($\alpha_g = 122 \pm 1°$, $\delta_g = 79 \pm 1°$ and , $\alpha_g = 124°$, $\delta_g = 80°$, respectively). Our result is also in good agreement with the geocentric radiant obtained by Jenniskens (2014) ($\alpha_g = 119.9 \pm 5.3°$, $\delta_g = 78.2 \pm 0.9°$). The averaged geocentric velocity calculated for these meteors was $V_g = 16.4 \pm 0.6$ km s$^{-1}$. This value is also in good agreement with the geocentric velocity of 15.86 km s$^{-1}$ predicted in Jenniskens (2006) and the 16.2 km s$^{-1}$ geocentric velocity predicted by Lyytinen (pers. comm.). The geocentric velocity observed by Jenniskens (2014) is $14.9 \pm 0.7$ km s$^{-1}$, which also fits fairly well our result. When the calculated orbital elements listed in Table 3 are compared to the orbit (J2000) of Comet 209P (a = 2.93 AU, e = 0.691, i = 19.30°, ω = 150.17°, Ω = 65.81713°, q = 0.9056 AU), the value of the Southworth and Hawkins $D_{SH}$ dissimilarity function (Southworth & Hawkins 1963) remains ≤ 0.15. This function measures the similarity of the orbit of the meteoroid and that of its parent body, with 0.15 being the usual cut-off value adopted to claim a positive association (Lindblad 1971a,b). A review about this topic can be found, e.g., in Williams (2011). The orbit of the comet was taken from the Jet Propulsion Laboratory (JPL) orbit database (http://neo.jpl.nasa.gov/).

The average tensile strength inferred for the Camelopardalids (($4.2 \pm 0.3$)·10$^4$ dyn cm$^{-2}$) is similar to the strength found for the Leonids (($6.0 \pm 3.0$)·10$^4$ dyn cm$^{-2}$) and the Orionids (($6.0 \pm 3.0$)·10$^4$ dyn cm$^{-2}$) (Trigo-Rodríguez & Llorca 2006, 2007). These values support that 209P/LINEAR meteoroids are very fragile and might be preferentially disrupted in the interplanetary medium in very short timescales compared with other meteoroid streams. A similar process was invoked to explain the decrease in the meteoroid flux in the optical range observed in 2011 October Draconids outburst (Trigo-Rodríguez et al. 2013). In any case, it seems that the decay of 209P/LINEAR meteoroids is quicker, perhaps by possible differences in the volume percent of material forming the matrix that glues the mineral grains and conform these cometary meteoroids.

The Na/Mg and Fe/Mg intensity ratios have been obtained from the measured relative intensities of the emission lines produced by Na I-1, Mg I-2 and Fe I-15. These ratios yield Na/Mg = 1.03 and Fe/Mg = 0.50. We have plotted these relative intensities in the ternary diagram shown in Figure 5. This plot shows that the point describing this spectrum corresponds to a non-chondritic particle, since this point deviates significantly from the expected relative intensity for chondritic meteoroids for a meteor velocity of about 18 km s$^{-1}$ (Borovička et al. 2005). Thus, the position of this experimental point in the diagram indicates that the meteoroid was depleted in Fe with respect to the chondritic value. We have recently discovered a similar bulk mineralogy in our recent paper on α-Capricornid meteoroids, associated with comet 169P/NEAT (Madiedo et al. 2014). Cometary particles are expected to be poor in metal grains, and pristine comets are expected to have a significant amount of organics and fine-grained Mg-rich silicates (Wooden et al. 2007). Then 209P aggregates may preferentially fragment in comet comae or during short stays in the interplanetary medium when they are released. We envision that the organic materials that might serve as the glue that binds aggregates of mineral subgrains desorb as previously envisioned (Jessberger et al. 2001; Wooden et al. 2007). This is also supported by looking at the quasi-continuous ablation behaviour observed in the videos of the meteors recorded by our CCD cameras (Figure 6). With respect to the spectrum of event SPMN220514B, we could only obtain the Na/Mg intensity ratio since the signal was not bright enough to show the contributions from Fe or any other elements. The value of this intensity ratio yields 1.07, which agrees with the Na/Mg intensity ratio obtained for the SPMN260514A spectrum.

It is interesting to note that complementary studies of this comet could give clues on the nature of 209P ejecta. We predict that the volatile component could be significantly depleted in volatiles or in fine-grained materials that, exposed to the interplanetary medium, are promoting an efficient disruption in the mineral grains that are forming pristine comets (Brownlee 2001, Brownlee et al. 2006).

## 5 CONCLUSIONS

We report accurate orbital and spectral information of meteoroids associated with a rare Earth encounter with 209P/LINEAR dust trails. Our CCD video cameras recorded Camelopardalid meteors from May 22 to May 28. The peak activity was much lower than expected, with a ZHR of around 20 meteors h$^{-1}$. Bright meteors were scarce, but we identified several bolides up to −8 absolute magnitude. Forward scatter data confirm that the maximum activity took place on 24 May 2014 between 6:00 and 8:00 UT, as predicted. These results are independently confirmed by the Canadian Meteor Orbit Radar (CMOR).

During the main activity period, the averaged geocentric radiant was located at $\alpha_g$ = 121.9 ± 1.1 °, $\delta_g$ = 78.3 ± 0.4 °, and the averaged geocentric velocity was $V_g$ = 16.4 ± 0.6 km s$^{-1}$. These values are based on the analysis of 7 double-station meteors imaged during the above mentioned time span and fit fairly well the predictions for these magnitudes.

The tensile strength of these meteoroids has been estimated and is consistent with fluffy aggregates of cometary origin, such as the Orionids and the Leonids. To reconcile the importance of the meteoroid flux detected by forward-scatter and radar techniques with the absence of optical meteors, we propose that the 209P/LINEAR meteoroids were preferentially fragmented during the short timescales in the interplanetary medium. On the other hand, it is also possible that the comet dust production was much lower than expected.

The analysis of the spectral data obtained for the Camelopardalids suggests that these meteoroids are non-chondritic. These particles exhibit a low abundance of Fe with respect to the chondritic value suggesting that the rock-forming materials in 209P/LINEAR are pristine.

## ACKNOWLEDGEMENTS

**Accepted for publication in Monthly Notices of the Royal Astronomical Society on Sept. 22, 2014.**

Meteor stations deployed at Sevilla, La Hita, Huelva, El Arenosillo, Sierra Nevada and La Pedriza have been funded by the first author. We acknowledge partial support from the Spanish Ministry of Science and Innovation (projects AYA2011-26522, AYA2012-31277 and AYA2012-30717). We thank the *AstroHita Foundation* for its continuous support in the operation of the meteor observing station located at La Hita Astronomical Observatory.

**REFERENCES**

Borovička J., Koten P., Spurny P., Boček J., Stork R., 2005, Icarus, 174, 15.

Brown P., 2014, Central Bureau Electronic Telegrams, 3886, 1.

Brownlee, D.E., 2001, In Accretion of Extraterrestrial Matter Throughout Earth's History, Ed. B. Peucker-Ehrenbrink and B. Schmitz, pp. 1-12.

Brownlee D.E., 2006, Science, 5806, 1711.

Ceplecha Z., 1987, Bull. Astron. Inst. Cz., 38, 222.

Jenniskens P., 2006, Meteor Showers and their Parent Comets. Cambridge University Press.

Jenniskens P., Lyytinen E., 2014, Central Bureau Electronic Telegrams, 3869, 1.

Jenniskens P., 2014, JIMO 42, 98.

Jessberger E.K., Stephan T., Rost D., et al. 2001, In Interplanetary Dust, E. Grün et al. eds., Springer-Verlag, Berlin, pp. 253-294.

Lindblad B.A., 1971a, Smiths. Contr. Astrophys., 12, 1.

Lindblad B.A., 1971b, Smiths. Contr. Astrophys., 12, 14.

Madiedo J.M., Trigo-Rodríguez J M., 2008, Earth Moon Planets, 102, 133.

Madiedo J M., Trigo-Rodríguez J M., Ortiz J.L., Morales N., 2010a, Advances in Astronomy, 2010, 1.

Madiedo J.M., Trigo-Rodríguez J.M., Lyytinen E. Data Reduction and Control Software for Meteor Observing Stations Based on CCD Video Systems. NASA/CP-2011-216469, 330, 2011.

Madiedo J.M., Trigo-Rodríguez J.M., Lyytinen E., Dergham J., Pujols P., Ortiz J.L., Cabrera J., 2013a, MNRAS, 431, 1678.

Madiedo J.M., Trigo-Rodríguez J.M., Konovalova N., Williams I.P., Castro-Tirado A.J., Ortiz J.L., Cabrera J, 2013b, MNRAS, 433, 571.

Madiedo J.M., Earth Planets and Space, 2014, 66,70.

Madiedo J.M., Trigo-Rodríguez J.M., Ortiz J.L., Castro-Tirado A.J., Cabrera-Caño J., 2014, Icarus, in press.

Maslov M., 2013, http://feraj.narod.ru/Radiants/Predictions/209p-ids2014eng.html

Moore C.E., 1945. In: A Multiplet Table of Astrophysical Interest. Princeton University Observatory, Princeton, NJ. Contribution No. 20.

Rao J., 2014, Sky & Telescope, 5, 30.

Southworth R.B., Hawkins G.S., 1963, Smithson Contr. Astrophys., 7, 261.

Trigo-Rodríguez J.M. and Llorca J., 2006, MNRAS, 372, 655.

Trigo-Rodríguez J.M., and Llorca J., 2007, MNRAS, 375, 415.

Trigo-Rodríguez J. M. et al., 2013, MNRAS, 433, 560.

U.S. Standard Atmosphere, 1976, NOA-NASA-USAF, Washington.

Wooden D., Desch S., Harker D., Gail H-P. and Keller L., 2007, In Protostars and Planets V, B. Reipurt et al. Eds., The Univ. Arizona Press, Tucson, USA, pp. 815-833.

Vaubaillon J., 2012,
http://www.imcce.fr/langues/en/ephemerides/phenomenes/meteor/DATABASE/209_LINEAR/2014/index.php

Williams I.P., 2011, A&G, 52, 2.20.

Ye Q, Wiegert P., 2014, MNRAS, 437, 3283.

**TABLES**

Table 1. Geographical coordinates of the meteor observing stations involved in this research.

| Station # | Station name | Longitude | Latitude (N) | Alt. (m) |
|---|---|---|---|---|
| 1 | Sevilla | 5º 58' 50" W | 37º 20' 46" | 28 |
| 2 | La Hita | 3º 11' 00" W | 39º 34' 06" | 674 |
| 3 | Huelva | 6º 56' 11" W | 37º 15' 10" | 25 |
| 4 | Sierra Nevada | 3º 23' 05" W | 37º 03' 51" | 2896 |
| 5 | El Arenosillo | 6º 43' 58" W | 37º 06' 16" | 40 |
| 6 | Villaverde del Ducado | 2º 29' 29" W | 41º 00' 04" | 1100 |
| 7 | Madrid-UCM | 3º 43' 34" W | 40º 27' 03" | 640 |
| 8 | La Pedriza | 3º 57´ 12" W | 37º 24´ 53" | 1030 |
| 9 | Folgueroles | 2º 19´ 33" E | 41º 56´ 31" | 580 |
| 10 | Montseny | 2º 32' 01" E | 41º 43' 47" | 194 |

Table 2. Trajectory and radiant data (J2000).

| SPMN code | Date and Time (UT) ±0.1s | M ±0.5 | $H_b$ (km) | $H_e$ (km) | $\alpha_g$ (º) | $\delta_g$ (º) | $V_\infty$ (km s$^{-1}$) | $V_g$ (km s$^{-1}$) | $V_h$ (km s$^{-1}$) |
|---|---|---|---|---|---|---|---|---|---|
| 220514A | May 22, 01h38m44.4s | -1.0 | 99.0 | 74.6 | 103.4±1.0 | 80.4±0.4 | 20.7±0.6 | 17.5±0.7 | 38.5±0.5 |
| 220514B | May 22, 22h02m39.2s | -3.5 | 95.2 | 79.0 | 102.9±3.0 | 82.7±0.2 | 20.4±0.7 | 17.1±0.8 | 37.9±0.6 |
| 230514A | May 23, 22h50m53.4s | 0.5 | 83.3 | 77.1 | 124.9±1.8 | 78.0±0.2 | 19.6±0.6 | 16.2±0.7 | 38.4±0.5 |
| 230514B | May 23, 23h13m36.1s | -1.5 | 93.1 | 76.2 | 125.2±1.4 | 77.4±0.3 | 19.3±0.4 | 15.8±0.5 | 38.2±0.3 |
| 240514A | May 24, 01h53m25.8s | 0.0 | 95.2 | 79.1 | 125.8±1.9 | 79.8±0.3 | 20.0±0.6 | 16.7±0.7 | 38.4±0.5 |
| 240514B | May 24, 02h07m10.7s | -0.5 | 91.8 | 77.2 | 118.6±0.8 | 78.6±0.3 | 19.9±0.5 | 16.5±0.6 | 38.4±0.4 |
| 240514C | May 24, 02h28m49.2s | -5.0 | 93.4 | 73.4 | 118.4±0.8 | 73.5±0.9 | 19.3±0.7 | 15.8±0.8 | 38.6±0.6 |
| 240514D | May 24, 03h51m23.3s | 1.5 | 92.7 | 75.2 | 121.1±0.4 | 80.2±0.4 | 20.1±0.5 | 16.7±0.6 | 38.2±0.4 |
| 240514E | May24, 04h23m33.1s | -0.5 | 94.9 | 78.4 | 119.0±0.5 | 80.7±0.3 | 20.9±0.3 | 17.0±0.4 | 38.3±0.3 |
| 260514A | May 26, 21h47m58.0s | -8.0 | 95.8 | 64.7 | 139.2±1.2 | 75.2±0.4 | 18.9±0.3 | 15.4±0.4 | 38.3±0.3 |
| 280514A | May 28, 00h06m48.7s | -4.0 | 97.3 | 74.1 | 100.6±1.9 | 83.3±0.5 | 22.7±0.7 | 19.8±0.8 | 38.9±0.5 |

Table 3. Orbital elements (J2000).

| Code | a (AU) | e | i (º) | Ω (º) | ω (º) | q (AU) |
|---|---|---|---|---|---|---|
| 220514A | 3.32±0.44 | 0.712±0.038 | 22.97±0.80 | 60.712581±10$^{-5}$ | 150.15±0.48 | 0.9563±0.0001 |
| 220514B | 2.80±0.40 | 0.657±0.049 | 23.47±0.88 | 61.52998±10$^{-5}$ | 150.10±0.82 | 0.9589±0.0006 |
| 230514A | 3.21±0.47 | 0.698±0.044 | 20.92±0.74 | 62.52292±10$^{-5}$ | 153.34±0.60 | 0.9683±0.0003 |
| 230514B | 3.08±0.29 | 0.686±0.029 | 20.37±0.51 | 62.53796±10$^{-5}$ | 153.36±0.44 | 0.9688±0.0005 |
| 240514A | 3.22±0.47 | 0.699±0.043 | 21.99±0.75 | 62.52316±10$^{-5}$ | 153.41±0.60 | 0.9685±0.0004 |
| 240514B | 3.17±0.33 | 0.696±0.032 | 21.44±0.69 | 62.65404±10$^{-5}$ | 151.82±0.36 | 0.9633±0.0002 |
| 240514C | 3.44±0.58 | 0.720±0.048 | 19.06±0.99 | 62.66791±10$^{-5}$ | 151.68±0.40 | 0.9618±0.0008 |
| 240514D | 3.10±0.29 | 0.688±0.030 | 22.19±0.72 | 62.72375±10$^{-5}$ | 152.31±0.22 | 0.9653±0.0005 |
| 240514E | 3.13±0.17 | 0.691±0.017 | 22.58±0.43 | 62.74463±10$^{-5}$ | 151.94±0.15 | 0.9639±0.0006 |
| 260514A | 3.14±0.25 | 0.689±0.024 | 19.54±0.38 | 65.36232±10$^{-5}$ | 155.67±0.45 | 0.9764±0.0009 |
| 280514A | 3.84±0.72 | 0.753±0.046 | 26.93±0.89 | 66.41621±10$^{-5}$ | 148.08±0.60 | 0.9474±0.0003 |

**FIGURES**

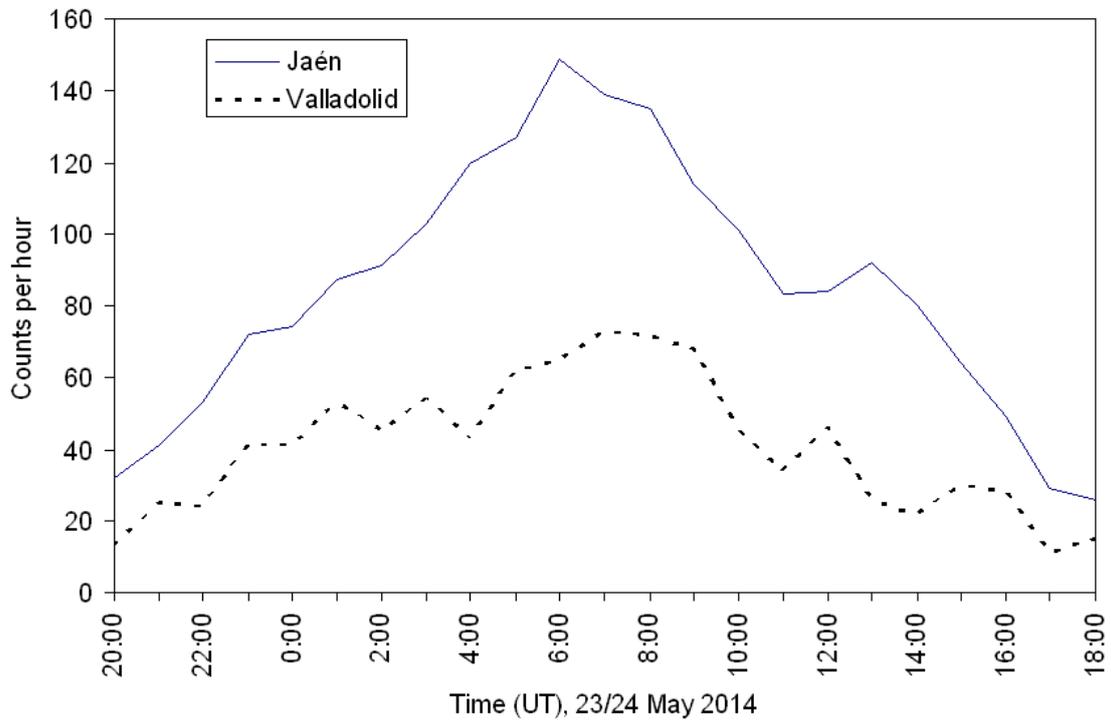

Figure 1. Activity level obtained by the forward scatter systems operating at the frequency of 143.05 MHz from Valladolid (Northwest of Spain) and Jaén (South of Spain).

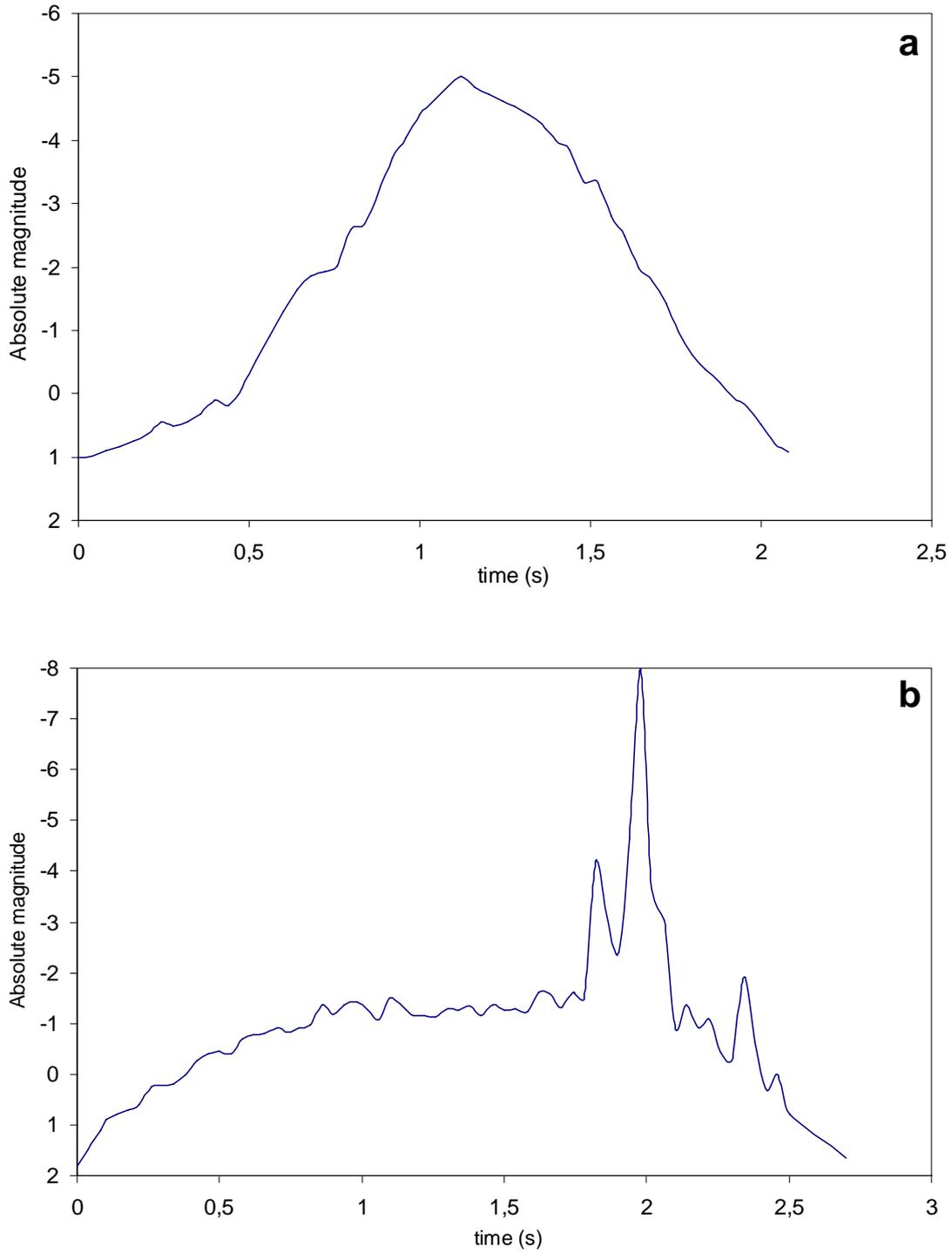

Figure 2. Lightcurve of the SPMN240514C (a) and SPMN260514A (b) Camelopardalid fireballs.

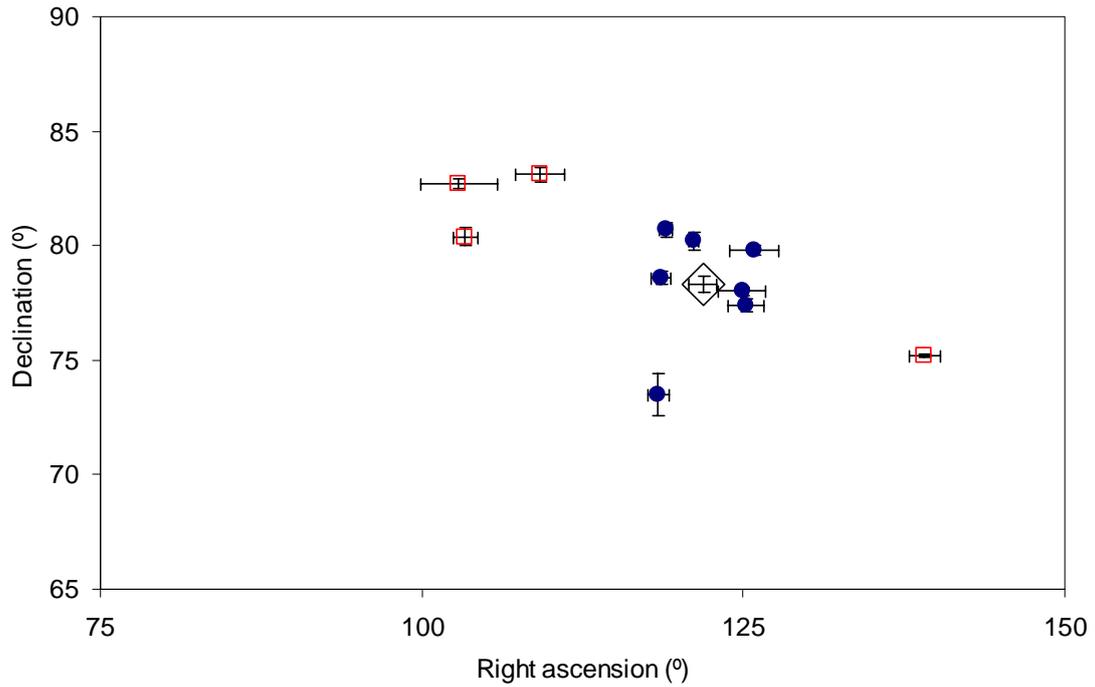

Figure 3. Geocentric radiant position for the Camelopardalid meteors discussed in this work. The full circles correspond to meteors observed during the shower main activity on May 23-24, and the open squares to meteors from the background component recorded outside that period. The open diamond shows the averaged radiant position during the main activity period.

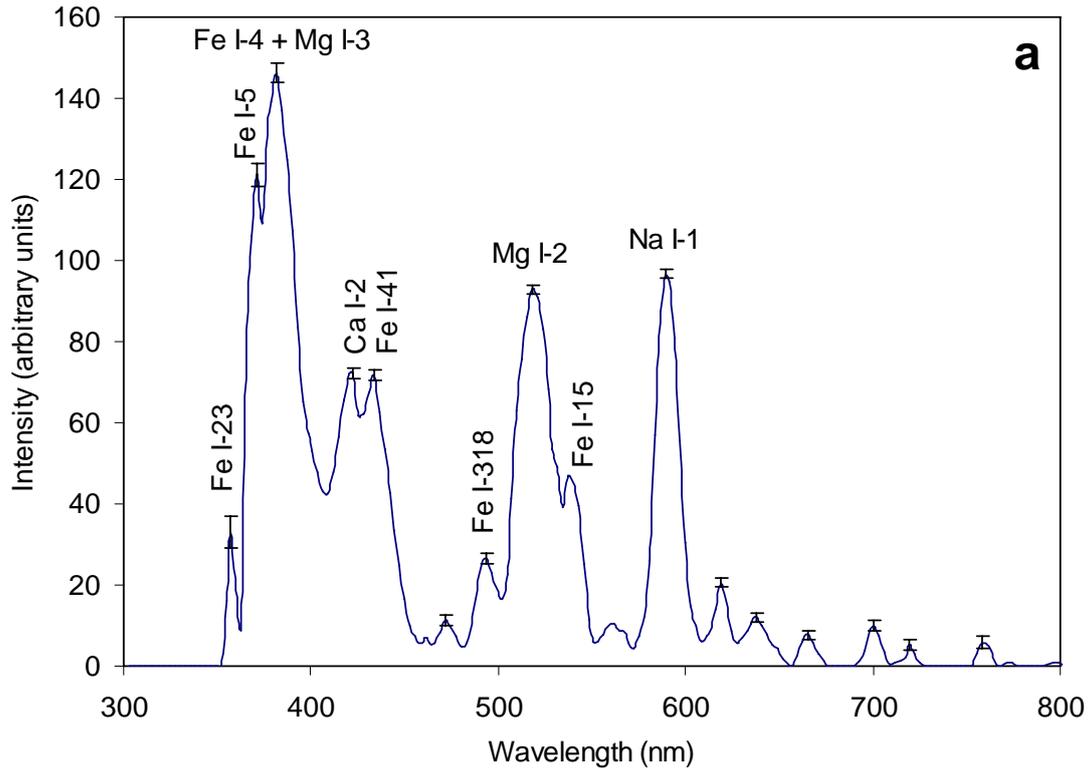

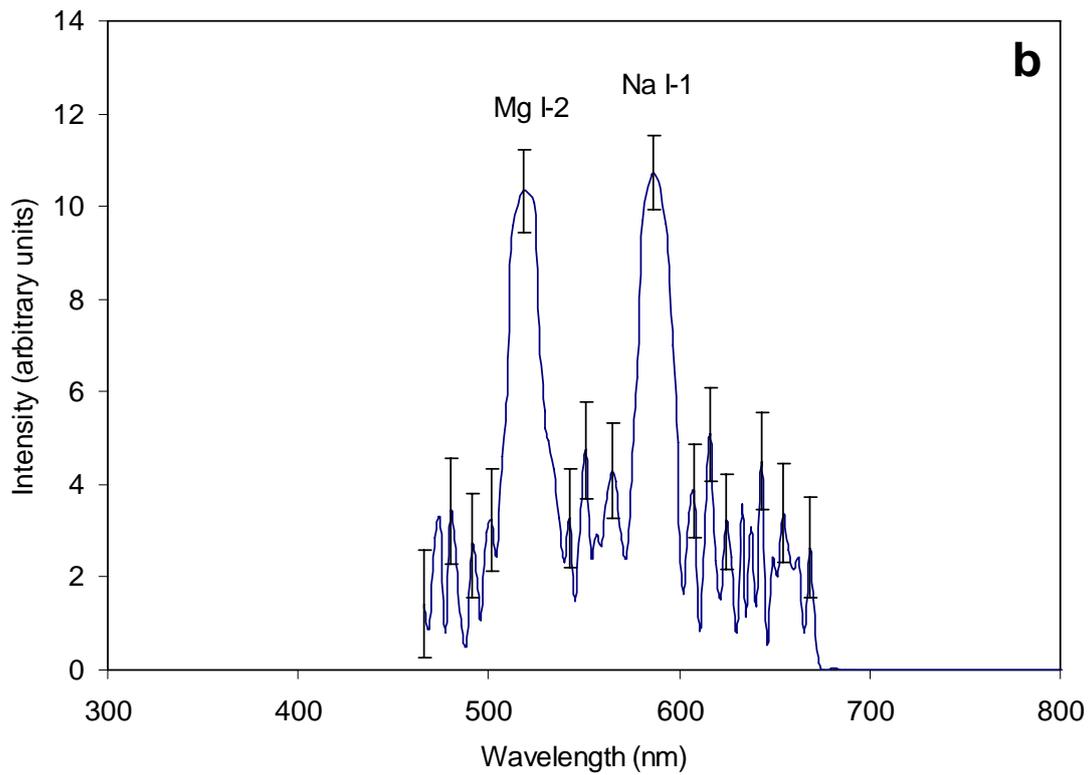

Figure 4. Calibrated emission spectrum of the SPMN260514A (a) and SPMN220514B (b) Camelopardalid fireballs, where the main contributions have been identified.

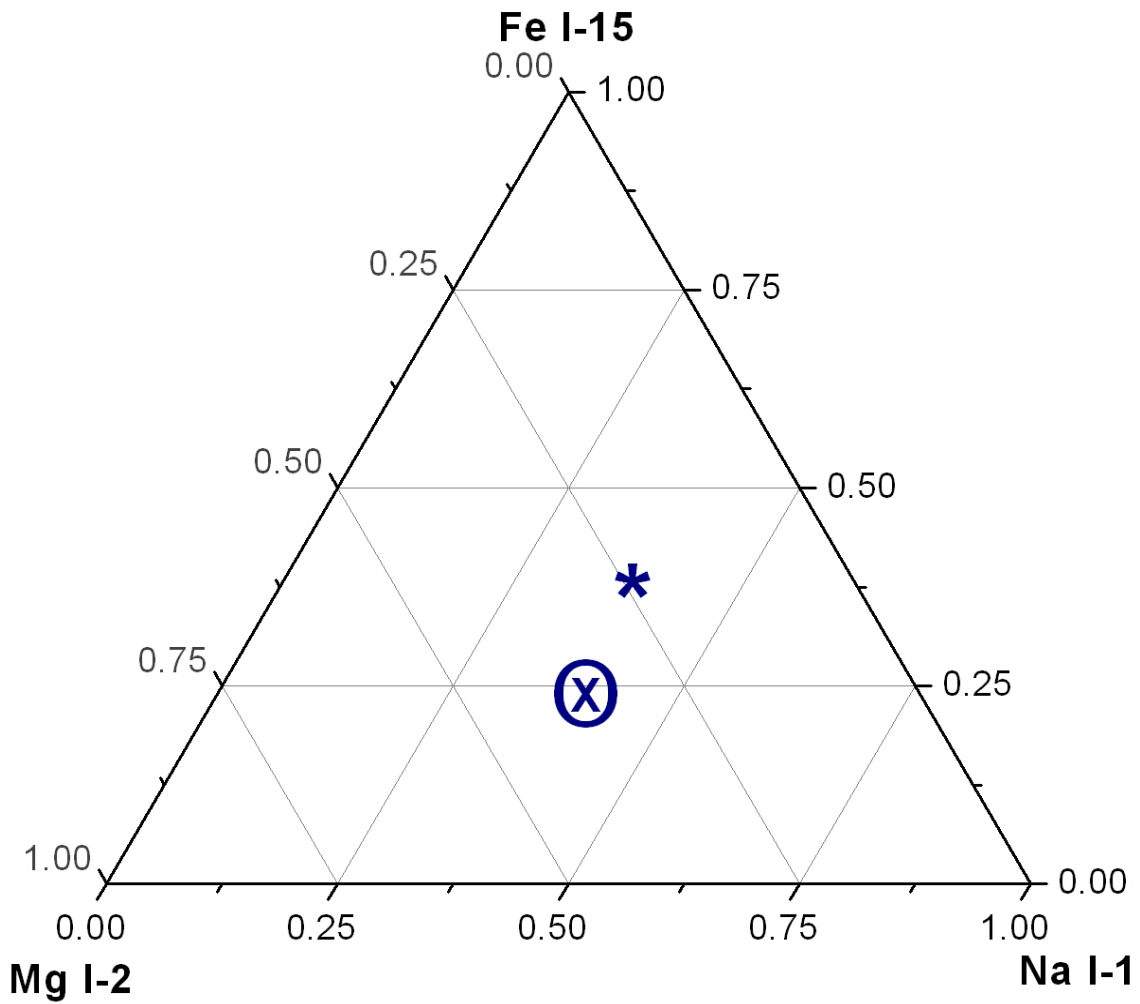

Figure 5. Asterisk: expected relative intensity of the Na I-1, Mg I-2 and Fe I-15 multiplets for chondritic meteoroids with a meteor velocity of 18 km s$^{-1}$ (Borovička et al. 2005). Cross: experimental result obtained for the SPMN260514A fireball. The area enclosed by the circle corresponds to the uncertainty (error bar) of this experimental value.

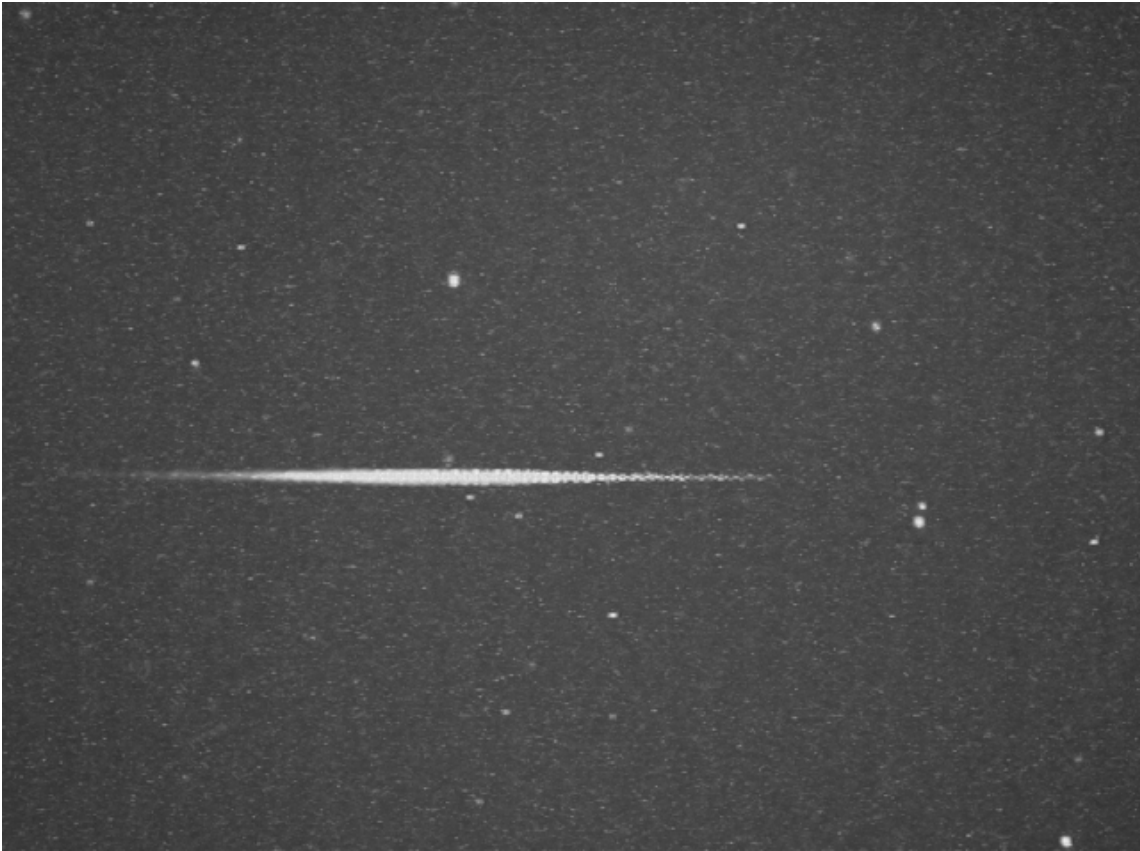

Figure 6. Sum pixel intensity image of the SPMN240514B event. This exemplifies the quasi-continuous ablation behaviour observed for Camelopardalid meteors (movie online).